# Imaging topological torus lattice from an electron crystal in twisted mono-bilayer graphene

Si-yu Li[1,3][†], Zhengwen Wang[1][†], Yucheng Xue[1], Yingbo Wang[1], Shihao Zhang[4], Jianpeng Liu[4], Zheng Zhu[5], Kenji Watanabe[6], Takashi Taniguchi[6], Hong-jun Gao[3,1][*], Yuhang Jiang[2][*] and Jinhai Mao[1][*]

[1]School of Physical Sciences, University of Chinese Academy of Sciences, Beijing, China

[2] Center of Materials Science and Optoelectronics Engineering, College of Materials Science and Optoelectronic Technology, University of Chinese Academy of Sciences, Beijing, China

[3]Institute of Physics, Chinese Academy of Sciences, Beijing, China

[4]School of Physical Science and Technology, ShanghaiTech University, Shanghai 200031, China

[5]Kavli Institute for Theoretical Sciences, University of Chinese Academy of Sciences, Beijing 100190, China

[6]National Institute for Materials Science, Tsukuba, Japan.

**A variety of exotic quantum phases of matter have been created by Van der Waals heterostructures. Moreover, these twisted heterostructures provide a feasible way of braiding correlation effect and nontrivial band topology together. Here, through a comprehensive spectrum study, we report the discovery of topological torus lattice in twisted mono-bilayer graphene. The strong Coulomb correlations give rise to an unusual charge localization behavior within the moiré supercell, leading to an electron crystal. The nontrivial band topology is encoded into the electron crystal, which would result in spatial modulated Chern numbers, and is evidenced by an emergent topological torus lattice state. Our result illustrates an efficient strategy for entwining and engineering topological physics with a strong electron correlation.**

Superposing two layers of van der Waals materials together has been an important thread for exploring the exotic quantum phenomena in condensed matter physics. One celebrated example is the realization of electronic flat bands near the Fermi energy ($E_F$) through a misalignment angle between the two layers, which provides a perfect platform for strongly correlated physics[1-5]. In these renormalized band structures by moiré superlattice, electron-electron interactions prevail over the almost vanishing kinetic energy of the non-dispersive flat bands, which endow the strong correlation effects[6,7]. Unconventional superconductivity[8,9], quantum anomalous Hall effect[10,11] and interaction-driven topological phases[12-14] have been reported from then on. While on the other side, the stacked van der Waals heterostructure could give rise to non-trivial topological Chern bands even without electron-electron interaction[15]. When the strong electron correlation and band topology are combined, unusual quantum phases are expected, and van der Waals heterostructures become the ideal platform for such a study. Here we report the correlation effect in twisted mono-bilayer graphene (tMBG) by scanning tunneling microscope (STM), where the strong correlation induced electron localization makes the band topology observable in real space without the necessity of any external magnetic field. As a result of such a subtle interplay between band topology and local correlation effects, spectrum imaging allows us to visualize an unprecedented topological torus lattice state in real space.

Figure1a shows a schematic atomic structure of the tMBG sample. The moiré superstructure comprises a triangular lattice of *ABB* regions (representing an on-top site stacking between the top monolayer and the bottom bilayer), which is alternatively surrounded by more structurally stable *ABA* and *ABC* regions. The tMBG sample is firstly scanned by STM in Fig.1b, where the twisted angle ($\theta = 1.04 \pm 0.02°$) could be directly deduced from the moiré periodicity $13.5 \pm 0.2$nm. The bright (dark, medium) spots here correspond to *ABB* (*ABA*, *ABC*) stacking regions in Fig.1a, analogous to the twisted double bilayer graphene (t-DBG)[16]. Band structure based on the tMBG at 1.04° is calculated by the continuum model without considering the electron-electron interactions (Fig.1c) (supplementary information, *SI*). The Dirac point ($V_D$) is flanked by two flat bands with weak dispersion in a quite narrow energy window, which allows for the realization of strongly correlated physics in such twisted samples[17,18]. Our calculation results also confirm the non-zero Chern number for the flat bands at this twisted angle (*SI*). To refrain from the electron-electron interaction, the dI/dV spectra (Fig.1d) in these three regions are collected at full filling, which all reveal the flat band materialized as two peaks around -60mV. Notice that, the flat band in tMBG are delocalized for backgate voltage at full filling, *i.e.*, have similar spectrum weight on *ABB*, *ABA*, and *ABC* areas around -60mV. This dI/dV spectrum is also consistent with our calculated local density of states, Fig.1e. However, when we flip to empty filling (Fig. 1f), the flat bands become quite localized on the *ABB* part, *i.e.*, with relatively large spectral weight in the *ABB* area compared to two other ones.

To scout for the correlated physics in tMBG system, Fig.2a to c collect dI/dV spectra from three high-symmetry areas (*ABB*, *ABA* and *ABC*) to show the evolution of those flat bands as a function of doping. Following such an evolution, we could deduce the global filling factor of the moiré unit cell $\nu = -4, -3 \ldots +3, +4$ corresponding to backgate voltage ($V_g$ -$V_{g0}$) of -28V, -14V…14V, 28V. On the hole-doped side ($\nu < 0$) where the flat band wavefunctions are mainly localized on the *ABB* site, we do not see any apparent band splitting in these spectra all over the three regions except the bandwidth of conductance flat band (CFB) getting broader. While on the electron-doping side, *i.e.*, $\nu > 0$, the gate-dependent dI/dV spectra of CFB present an intriguing mode. For the *ABB* and *ABA* stacking regions (Fig. 2a and b), there is no apparent peak-splitting behavior under all fillings but a modulation of the energy difference between valence flat band (VFB) and CFB. But for *ABC* region (Fig. 2c), once the CFB touches $E_F$, the single band splits into two branches and with a peak-to-peak energy difference, $\Delta E \sim 10$meV (dashed square in Fig. 2c). We attribute the emergence of this correlated state here to the relatively narrower bandwidth of CFB on the electron-doped side, $w_{CFB} = 20$meV, as compared to the Coulomb interactions, $U = \frac{e^2}{4\pi\varepsilon_0\varepsilon_r L} \sim 42 meV$ that directly estimated from the moiré superlattice periodicity $L$ and dielectric constant of the environment $\varepsilon_r = \frac{\varepsilon_{BN}+1}{2} \sim 2.5$. The smaller $w_{CFB}/U \sim 1/2$ meets the threshold to active a strongly correlated insulating state[19]. Interestingly, simultaneous to the CFB splitting at these partial fillings, the lower VFB also splits even though it is already filled and far away from the Fermi level. The peak splitting behavior could not be explained by a single-particle band picture, so further corroborate the correlated physics[20]. It is also worth noticing that even though the strong correlation effect affects both CFB and VFB in the *ABC* regions, the spectra in *ABB* and *ABA* areas are still keeping a non-splitting status, *i.e.*, the splitting induced insulating phase is only confined within the *ABC* regions. The local band splitting behavior in *ABC* but absent in both *ABB* and *ABA* suggests that we should have a spatial modulated electron-electron interaction strength that can be potentially described by an emergent real-space lattice model with localized interactions. This spatial variation of electron correlation is out of the expectation from a single-particle picture or a mean-field description of long-range Coulomb interactions and should be a result of a strong local correlation effect.

Interestingly, the CFB for these three regimes crosses $E_F$ at different global fillings, resulting in a spatial varied local filling. This suggests that the system would incline a spatial inhomogeneous charge redistribution[21]. The spatial variation of flat band energy at a fixed global filling, *e.g.* $\nu = 3$, further supports this perspective (Fig. 2d). An obvious band bending near the *ABB* region shows up (indicated by the arrows), which agrees with an additional electron carriers' existence.

An elaborate inspection on those charge density modulation allows us to extract a real space cascade filling of the moiré unit cell among the three high symmetric regions. In Fig.3a-b, we show the deduced filling sequence of the moiré unit cell directly from Fig.2a-c and conceive the electron crystalline structures as a function of filling factors. At $\nu = 1$, the electrons start to occupy both *ABB* and *ABC* regions. The *ABB* regions are fulfilled (red circles), however, *ABC* regions are only half-filled due to its band splitting (gray circles). At $\nu = 2$, the *ABB* and *ABC* regions keep the same local filling ratios, and the *ABA* region starts to be partially filled by electrons. For $\nu = 3$, both *ABB* and *ABA* regions are full filled, while the *ABC* regions are kept half-filling. Following the above observation, we conclude that there is an aggregation priority for doping electrons among the three areas. It is in contrast to the delocalized flat band wavefunctions distribution in the real space, so this electron aggregation is a kind of transition from a uniform electron gas into a localized crystal lattice, *i.e.*, an electron crystal formation. The electron correlation plays an indispensable role in this spontaneous electron crystal formation. To confirm that, we examine the sample at a non-magic angle but failed to see such charge modulation (*SI*). To catch this electron crystallization process, in supplementary materials, we provide a toy model with a tripartite structure representing *ABB*, *ABA*, and *ABC* sublattices. Relying on our observed different spectrum splitting behaviors, in this model we set three local Coulomb interaction strengths for each sublattice site. The model could qualitatively repeat the real space charge density modulation (*SI*). Based on these understandings, we try to visualize this electron crystal in real space. Fig. 3c-f collect dI/dV maps near $E_F$ and calculate the local filling ratio $R$ of CFB at each point. Here we define the local filling ratio as $R = A_{LB}/(A_{UB} + A_{LB})$ by following our previous method in twisted bilayer graphene[3] to unveil the charge ordering, and $A_{LB}$ ($A_{UB}$) is the area under the dI/dV spectrum for CFB below (above) $E_F$ part after subtracting the spectrum background (*SI*). Due to the spatial charge density modulation, we should expect a spatial modulated local filling ratio $R$ to reflect the electron crystallization. Fig. 3c-f shows the color contour of extracted local filling ratio at fixed global filling ν, where the results both qualitatively and quantitatively fit the deduced electron crystal in Fig.3b. Initially for $ν = 0$, there are no electrons in the whole sample, leading the local filling ratio all around zero everywhere (dark blue color in Fig. 3c). At $ν = 1$ (Fig. 3d), *ABB* regions first collect most of the doped electrons (red triangle areas). Combined with half-filled *ABC* regions, they do show a honeycomb electron lattice which coincides with the prediction picture in Fig. 3b. As the filling factor continues to increase (Fig. 3e-f), *ABA* regions are the second to realize full filling (central red ball of the dashed structure in Fig. 3f). In the range of $ν = 1, 2, 3$, *ABC* region always keeps partially filled, and the CFB band keeps splitting with a correlated gap. Until the global filling factor reaches 4, the local filling ratio also gets fully filled in the whole sample (bare red contour in the inset of Fig. 3f). Following this method, we directly visualize the electron crystal in tMBG formed at partially global fillings. The electron crystal structure extracted here corroborates the charge density modulation and crystallization near the integer fillings.

Real-space cascade filling among the three high-symmetry areas makes the electron crystal here different from a charge density wave (CDW) or recent reported Wigner crystal in twisted transition metal dichloride[22,23]. First, it gives a dramatic modulation on the local band filling, empty in *ABA* and full in *ABB* at $ν = 1$, in strong contrast to a slight charge modulation in CDW physics. Second, different from the CDW and Wigner crystal, our charge density modulation does not break either translational or rotational symmetry and even does not need to enlarge the unit cell as compared to those physics in moiré superlattice[22,24,25]. Third, even though charge modulation forms, there is no global gap opening to save the total energy like in CDW[26]. Moreover, in contrast to the Wigner crystal that is limited to the low carrier concentration, our electron crystal could happen within most of the partial filling range, like $\nu = 1, 2, 3$. Those features identify our charge density modulation as a peculiar format of electron crystal.

Strong electron-electron interaction does not only promote an electron crystal formation but also provides a new form of realizing correlated insulating states. As mentioned above, during the electron crystal formation process, the local filling of CFB also gets modulated. The collaboration of charge density modulation by the electron crystal with the local band filling gives a global metal-insulator transition. For example, at the filling factor $\nu = 1$, the CFB at *ABB* part is full filled, whereas still keeps empty at *ABA*. At the same time, the flat bands at the *ABC* regions split with the lower (upper) branch occupied (empty). All those three areas are insulated due to the vanishing density of states at $E_F$, and when knitted together would make a global insulating phase (see *SI* for other fillings). This is quite different from other twisted systems with flat band splitting to open a gap for the insulating phase at the integer fillings. The alternate filling in real-space moiré unit cell provides a new mechanism for realizing the correlated insulated state at integer fillings. Such a real-space picture for the formation of correlated insulator state is especially unusual for twisted graphene systems, in which the nonzero Chern numbers impose obstructions in constructing exponentially localized Wannier functions in real space, which naturally impedes a real-space description to the strongly correlated problem.

The special route on realizing the correlated insulating state gives us a hint on the mechanism for the electron crystal formation. Despite the topological obstructions, we suppose the system could still be described by a real-space lattice model in some way with local interactions. Then in a homogenous phase, it is hard to form a Mott-like insulator as the doping deviates from the one-half. But when the intersite interaction is sufficiently large, the system could roll into a charge ordering phase with one sublattice gets closer to half-filling and the other keeps empty. This lattice analog of electron crystal could help the realization of the insulating state. Here in this tMBG system, as mentioned in Fig. 1, the electronic states first give a quite uniform distribution. The comparable valence bandwidth to the Coulomb interaction strength further facilitates the charge separation as predicted for the strongly correlated electrons in the two-band model. This may be a motivation why the electron crystal is preferred here[27].

Entwining the electron crystal with the local flat band filling modulation does not only offer a way of realizing the correlated insulating state but also enables to directly study the nontrivial band topology without any external magnetic field. Due to the relatively low crystal symmetry, flat bands in tMBG could carry a nonzero Chern number (*SI*). In our calculations, the conduction flat band in tMBG at $\theta = 1.04 \pm 0.02°$ do have a finite Chern number $C = \pm 1$ for $\pm K$ valley. In a simplified picture, the system's Chern number is determined by the filling status of the flat bands. We consider three possible situations of flat band fillings, which exhibit different Chern numbers or valley Chern numbers. First, when two flat bands (per spin) are empty, the Chern number contributed by the CFBs is zero; second, if the two flat bands from the two valleys (for either one spin species or both spin species) are fully occupied, there would be nonzero and valley-contrasting Chern numbers from the CFBs although the total Chern number is still zero; lastly, if only the flat band from one valley is occupied, which is split from the other valley, the net Chern number from the CFBs would be nonzero for our tMBG system. The electron crystal discussed above gives a spatially modulated flat band filling, which means it could induce a real space Chern number variation within the moiré cell, which can be mathematically described by the local Chern marker[28]. As a result, topologically protected interface states would be expected at the boundary of two areas with different local Chern numbers[29,30]. The periodic real-space modulation of the Chern numbers would give rise to a periodic lattice of the gapless topological interface states, forming torus ring in real space, thus we call such a state a topological torus lattice state.

To search for those expected topological metallic states given by the electron crystal, we perform the dI/dV mapping near $E_F$ at $\nu = 1$ (Fig. 4). Around this filling factor, the CFB in *ABB* regions are full filled which have nonzero valley Chern numbers, while *ABA* regions are empty, so the *ABB* sites should be topological nontrivial and *ABA* sites act as a vacuum there. At the boundary, an enhancement of the density of states should be expected due to the gapless topological states. Experimentally, we do observe a succession of torus structures enclosing the *ABA* sites, serving as strong evidence for the expected topological states (Fig. 4a). The density of state along a line-cut across the torus does show an enhancement in Fig. 4b, which also agrees with the scenario of topologically protected interface states. In order to confirm the topological origin of the torus states, we perform the same experiments on the other samples without strong electron correlation nor real-space electron aggregation, such that there is no real-space modulation of local Chern numbers, and we do not observe the torus structures (*SI*). Therefore, we conclude that the strong electron correlation, and the real-space charge density modulation, as well as the nontrivial band topology, together play crucial roles in forming those topological torus states (*SI*). Importantly, the radius of the torus structure does not disperse with the energy, or doping variation under external field, but is pinned at the boundary (Fig. 4c-d). However, when we tune the filling to full filling, all those torus structures disappear (*SI*). We also exclude other possibilities for generating the torus structures, which are presented in detail in *SI*. Taking all these observations into account, it is legitimate to claim that the torus structure originates from the topologically protected boundary states between the different regions in the moiré cell. The combination of the correlation-driven electron crystal with band topology allows for the formation of a periodic torus structure and makes a lattice of topologically protected states. Since the topological torus structure is based on the moiré structure, which means we may have the possibility to engine the topological states easily by strain besides the electric or magnetic field.

The topological torus state forms a closed circle, which means that the topologically protected state also emerges between *ABA* and *ABC* regions. Even though the strong electron correlation can split and renormalized the flat bands such that the original single-particle band topology may not be simply applied, the survival of this torus state requires that the *ABC* region is also topologically nontrivial. While the absence of a topologically protected state at the boundary between *ABB* and *ABC* further indicates the topological phases in those two areas should be equivalent. This implies that the electron correlation in *ABC* region has to give the same Chern number as that in *ABB*. For example, this could happen when the CFBs are spin split in the ABC region, and in the meanwhile, the correlation effects also locally change the valley Chern numbers of the spin-polarized flat band such that they are the same as those in *ABB* without spin splitting. A more quantitative theoretical modeling of such topological torus state is beyond the scope of the present work, which we leave for future study. The band topology and strong correlation-driven topological Chern band together give the torus structure. In recent twisted graphite systems, the strong correlation has been a powerful tool for driven the nontrivial topological phase. Our electron crystal demonstrates a new vehicle on coupling those topological states from weak correlation and strong correlation and even could braid them together for novel topological phase.

# Figures

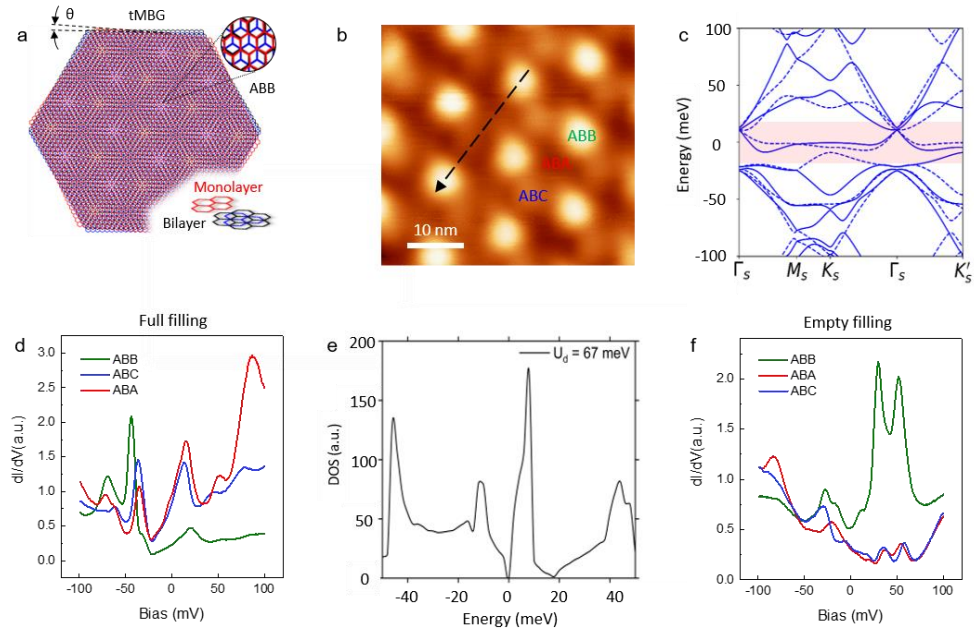

**Fig.1 Electronic structures of the twisted mono-bilayer graphene (tMBG) at 1.04°. a**, A schematic atomic structure of the tMBG. θ is the twisted angle between the top monolayer (red honeycomb structure) and the bottom bilayer graphene (blue and black structure). **b**, STM topography of tMBG that showing the moiré patterns ($V_b$ = -1V, I = 50pA). According to the collected dI/dV spectra, different stacking configurations are assigned, and the high-symmetry regimes are labeled by *ABB*, *ABA*, and *ABC* according to the atomic registry. The dashed line indicates the trace for the spatial evolution of flat band in Fig. 2**d**. **c**, The calculated band structures for tMBG by the continuum model. Red shadow highlights the flat band. **d**, dI/dV spectra for the three high symmetry regimes that are taken at full-filling state ($V_g$ = +40V) ($V_b$ = -100mV, I = 200pA). **e**, The DOS that directly deduced from the band structure in **c**. The Fermi level here corresponds to the Dirac Point in experimental data. **f**, dI/dV spectra for the three high symmetry regimes that are taken at empty filling state ($V_g$ = -40V) ($V_b$ = -100mV, I = 200pA).

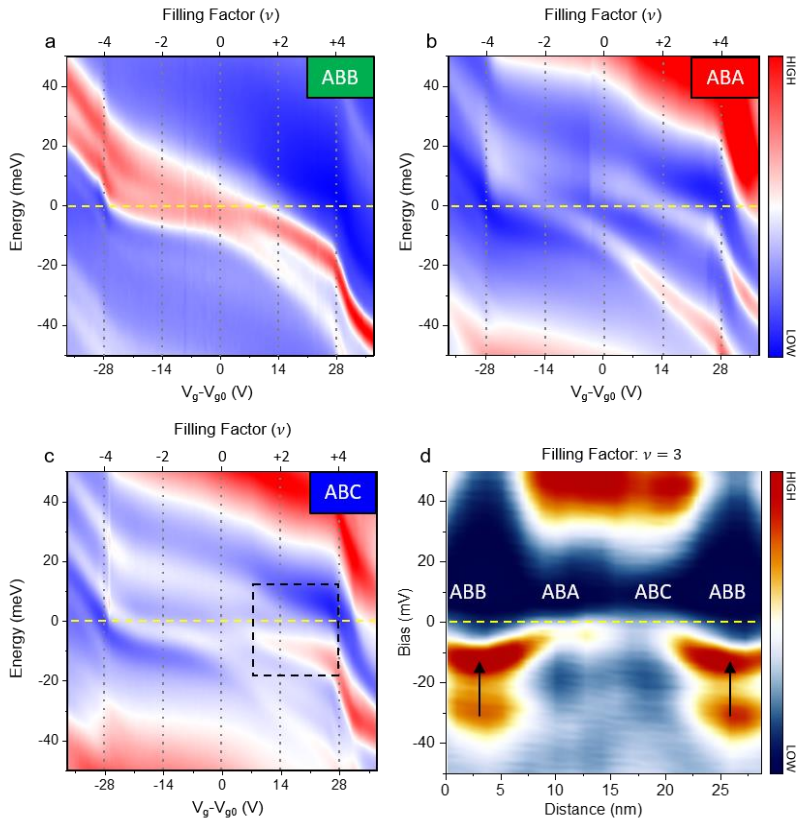

**Fig.2 Charge separation**. **a-c**, The color contour of the dI/dV evolutions as a function of doping level ($V_g$) for *ABB*, *ABA* and *ABC* respectively ($V_b$ = -50mV, I = 200pA), here $V_{g0}$ = 3V is the gate voltage to make the system charge neutral. The filling factors ν that are deduced from the gate voltages are also labeled. The yellow dashed lines in each panel indicate the Fermi level that we set as $E_F$ = 0meV. The dashed square in **c** highlights the correlation gap in *ABC* region around Fermi level. **d**, Line profile of dI/dV spectra to show the spatial evolution of flat band for ν = 3 along the dashed line in Fig.1b. The arrows show the electron pools in *ABB* region.

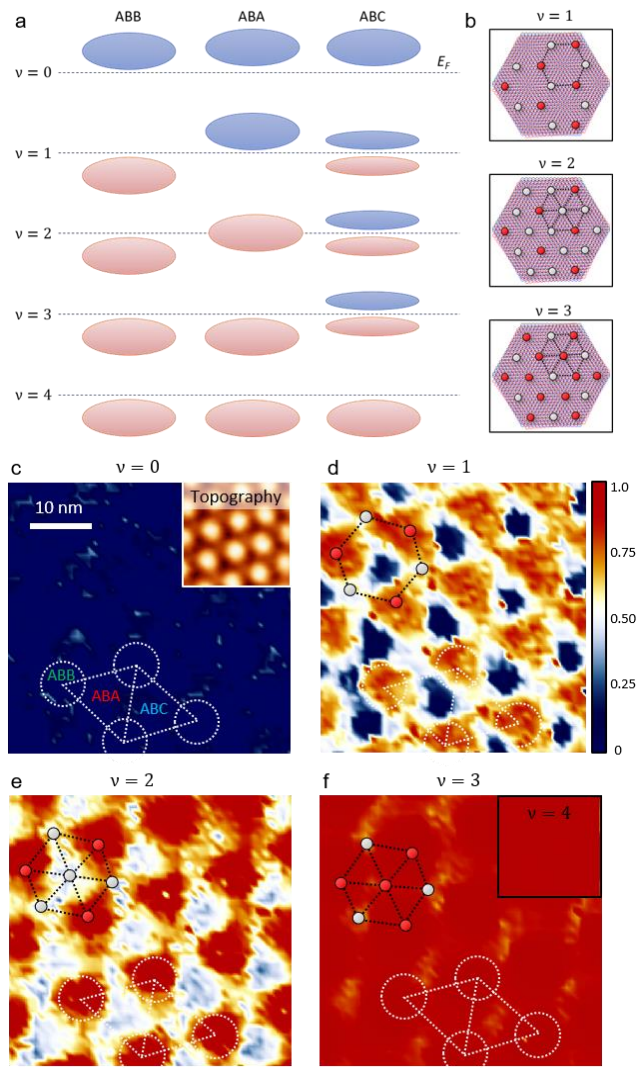

**Fig.3 Electron crystal. a**, Schematic drawing shows filling sequences of *ABB*, *ABA*, and *ABC* regions as a function of global fillings ν that is listed on left. The dashed lines indicate the Fermi level under each filling factor. The red (blue) oval highlights the partial flat bands with (without) electrons occupation. **b**, Proposed lattice structure of the electron crystal. Red (gray) circles superposed on three high-symmetry regions indicate the positions fully (partially) occupied by electrons. **c-f,** Extracted local filling ratio from dI/dV maps for different filling factors ν = 0 (**c**), ν = 1 (**d**), ν = 2 (**e**), ν = 3 (**f**), and ν = 4 (inset of **f**). ($V_b$ = -100mV, I = 200pA). The inset in **c** shows topography simultaneously obtained in these maps, which can offer the structure of the moiré superlattice and the position of three high-symmetry regions (white dotted lines in **c-f**). The red (gray) circles superposed on these maps represent areas with full (partial) filling, which form the similar superstructures with those in **b**. The color bar shows the local filling ratio (R) from 0 to 1.0.

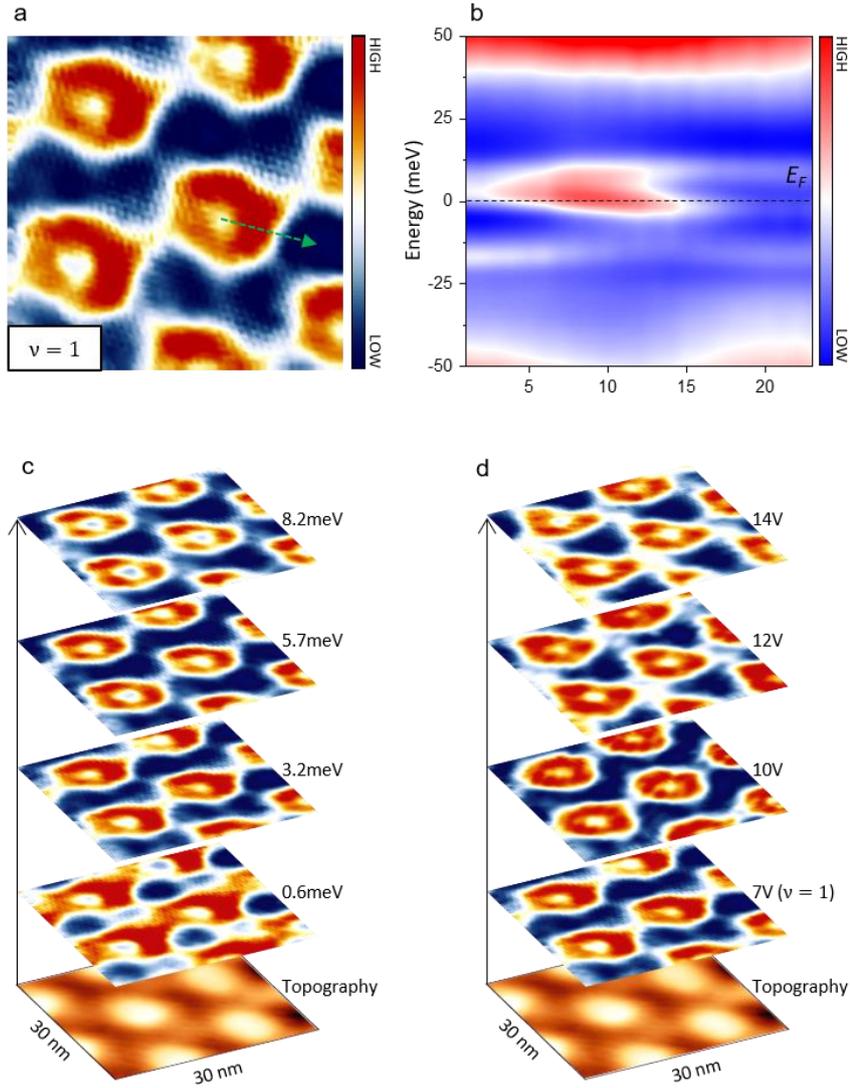

**Fig. 4 Topological torus lattice. a**, A typical dI/dV map at ν = 1 near $E_F$. The torus structure show up with bright red color ($V_b$ = -50mV, I = 200pA). **b**, Evolution of the dI/dV intensity from the *ABA* to *ABC* region along the arrow that crosses the torus in **a**. The dashed line indicates Fermi Level ($E_F$). The red color near $E_F$ at the torus position indicates the enhancement of LDOS. **c**, Stacking of dI/dV maps at ν = 1 as a function of energy which are labeled nearby ($V_b$ = -50mV, I = 200pA). The bottom image shows the STM topography simultaneously obtained in these maps. **d**, Stacking of dI/dV maps as a function of doping level which are labeled nearby ($V_b$ = -50mV, I = 200pA). The bottom image shows the STM topography simultaneously obtained in these maps.